# Field deployment of a laser wakefield accelerator for on-site application

Bo Guo[1], Xiaonan Ning[1], Dexiang Liu[2], Yue Ma[3], Weiwang Zeng[2], Mingyuan Wei[2], Shengtai Wei[2], Jianfei Hua[2], Yang Wan[4], Wei Lu[3,1,2]*

[1]Beijing Academy of Quantum Information Sciences; Beijing 100193, China.

[2]Department of Engineering Physics, Tsinghua University; Beijing 100084, China.

[3]Institute of High Energy Physics, Chinese Academy of Sciences; Beijing 100049, China.

[4]Laboratory of Zhongyuan Light, School of Physics, Zhengzhou University; Zhengzhou, 450001, China.

*Corresponding author. Email: weilu@ihep.ac.cn

**Successive innovations in particle accelerators have continually expanded the frontiers of scientific discovery. Laser wakefield accelerators promise to transform science, medicine, and industry, yet moving them from laboratory demonstrations to reliable real-world operation has remained a central, long-standing challenge. Here we report a field-deployable system that produced 100-MeV-class electron beams with 1%-level energy stability during 72 hours of continuous operation and supported routine full-power use throughout a seven-month field trial in an industrial setting. Applied to in situ micro–nondestructive testing, the system generated tens-of-MeV bremsstrahlung X-rays that enabled three-dimensional microtomography of dense materials at sub-50-μm spatial resolution and revealed 100-μm-scale internal defects in large composite structures, extending the capabilities beyond those of existing high-energy X-ray sources. These results mark a transition of laser wakefield acceleration from laboratory proof of concept toward practical deployment in scientific and industrial applications.**

Advances in particle accelerator technology have underpinned many of the most transformative scientific discoveries and technological advances, from revealing fundamental particles to resolving the structures and dynamics of complex matter. Since its conception in 1979(*1*), laser wakefield acceleration (LWFA) has emerged as a disruptive complement to conventional radiofrequency (RF) accelerators(*2-4*), with the long-standing goal of bringing compact, high-quality particle and radiation sources beyond specialized laboratories into practical use. In an LWFA, an ultrashort, high-intensity laser pulse ionizes a gaseous medium to drive a trailing plasma wave that sustains accelerating gradients exceeding 100 GV/m(*5-9*)—more than three orders of magnitude higher than those available in RF cavities—offering a viable pathway to compact, high-quality particle sources(*10-12*). Owing to their micrometer-scale accelerating structures, LWFAs naturally produce relativistic electron beams with femtosecond durations(*13, 14*) and micrometer transverse sizes(*15, 16*), enabling investigation of matter and dynamics under extreme conditions with unprecedented spatiotemporal resolution.

Over the past two decades, substantial efforts have been focused on improving beam quality(*17-22*) for prospective applications in next-generation electron-positron colliders and X-ray free-electron lasers(*23, 24*). In parallel, motivated by the demand for compact, high-performance radiation sources(*25-28*) in biology, medicine and materials science, laboratory-scale demonstrations of LWFA-based applications—including micro-X-ray imaging(*29-33*), ultrafast spectroscopy(*34, 35*) and radiotherapy(*36*)—have advanced steadily. Despite continuing improvements in stability(*37, 38*), these milestones remained predominantly proof-of-principle demonstrations in tightly controlled laboratories, and no decisive demonstration established reliable operation under real application constraints. The considerable footprint and environmental sensitivity of existing systems have therefore sustained long-standing doubts regarding their real-world applicability, particularly for in situ deployment.

Here we demonstrate a compact, robust LWFA system that crosses this critical boundary by sustaining operation in an uncontrolled industrial environment. Throughout a seven-month trial, the system maintained full-power operation for an average of 10 hours per day, including uninterrupted runs exceeding 72 hours with a root-mean-square (r.m.s.) electron-beam energy stability of 1.3% (shot-to-shot). To meet growing demand for fine-structure characterization in advanced technologies, we deployed LWFA-driven tens-of-MeV bremsstrahlung X-rays in two challenging micro-nondestructive testing (μNDT) scenarios: the inspection of dense metals and the evaluation of large, complex composites. The X-ray source, featuring a 38-μm source size, operated stably for continuous runs exceeding 10 hours in a conventional industrial NDT workshop. Using more than 60,000 consecutive exposures, we performed micro-computed tomography (μCT) on a nickel-alloy aeroengine turbine blade with a spatial resolution of 43 μm, unlocking capability of sub-50 μm tomography for tens-of-MeV high energy X-rays regime. By combining image stitching with turntable-offset scanning, the effective field of view (FOV) was expanded to > 650 mm. A subsequent μCT scan of a 640-mm aluminum-ceramic phantom, reconstructed from more than 170,000 consecutive exposures, clearly resolved internal defects on the 100-μm scale. These results mark a milestone for the field deployability of LWFA and establish its transition into a reliable tool for high-precision in situ applications.

## Overview of the field-deployable LWFA system

Figure 1A illustrates the schematic layout of the entire system installed in an industrial X-ray NDT workshop lacking environmental controls. The LWFA is housed within a shipping container measuring 7,450 mm × 2,450 mm × 2,800 mm, enabling rapid deployment across diverse field environments (Fig. S1). Although only a minimal environmental conditioning is provided by the container (ISO Class 7 cleanliness, temperature stability of ± 1 °C, and basic humidity control < 50%), the system maintains long-term stable operation through the refined environmental control integrated locally within each module. Inside the container, the main assembly occupies a footprint of 3,500 mm × 1,500 mm, comprising a compact, self-developed 40 TW Ti:Sapphire laser system coupled to vacuum chambers.

Following the final-stage amplifier, the 800-nm laser pulses are compressed to 25 fs (full width at half maximum, FWHM) by a pair of diffraction gratings within a compressor chamber. A nitrocellulose membrane isolates the compressor from downstream components to preserve the vacuum quality. The compressed pulses are then focused onto a helium-nitrogen (99 : 1) gas mixture ejected from a de Laval nozzle to drive the laser wakefield acceleration in the LWFA chamber. After exiting the gas jet, the electron beams are either characterized by a magnetic spectrometer or directed onto a tungsten target to produce bremsstrahlung X-rays (Fig. 1B).

To optimize the laser-to-electron energy conversion efficiency, wavefront aberrations of laser beams were minimized by adjusting the lenses and off-axis parabolic mirror, ensuring that 45% of the pulse energy was enclosed within the FWHM focal spot. Gas target parameters, including nitrogen concentration and plasma density, were finely tuned to match the laser pulse energy and focal position. Stable quasi-monoenergetic electron beams were generated via ionization-induced injection(*39, 40*) with tunable mean energies of approximately 70-100 MeV, total charges up to ~1 nC, and repetition rates up to 10 Hz (Fig. S2). Full-power operational statistics over a seven-month trial are summarized in Fig. S1D, showing an overall availability exceeding 70% of the total days. On average, the system maintained full-power operation in electron / X-ray mode for more than 10 hours per day. The longest continuous run reached 12 days, sustaining over 13 hours of operation daily.

For real-world applications, superior long-term, continuous operational reliability is paramount. To evaluate this performance, electron-beam stability was assessed by continuous spectrum acquisition over a 72-hour period at a repetition rate of 0.1 Hz (Fig. 2A). Despite ambient temperature variations outside the container ranging from 24 °C to 29 °C, the LWFA sustained stable operation. Figure 2B summarizes the corresponding statistics of mean energy and charge (> 70 MeV), revealing r.m.s. shot-to-shot fluctuations of 1.3% in mean energy (86.2 ± 1.1 MeV) and 9.8% in charge (164 ± 16 pC). Compared to shot-to-shot stability, integrated stability is often more critical for practical applications. For example, high-energy X-ray three-dimensional (3D) μCT scans, whether driven by LWFAs or conventional micro-focus RF systems, typically require several hours of continuous acquisition, with individual projections often taking minute-scale exposure times. Over the 72-hour trial, the integrated stability over 10-minute intervals (corresponding to 60 consecutive pulses) improves to 0.4% in mean energy and 1.9% in charge.

## μCT of dense industrial components

By directing LWFA-accelerated electron beams onto a tungsten target, bremsstrahlung X-rays in the tens-of-MeV range were generated. The X-rays exhibited stability comparable to that of the electron beams. Figure 3A shows the radiation dose stability of 1,000 consecutive X-ray pulses measured 1 m downstream from the source at 3.3 Hz. The average single-shot dose was determined to be 157 ± 21 μGy, exhibiting a 100-shot dose stability of 1.4% (r.m.s.). This yields an average dose rate of ~ 3.1 cGy/min @ 1m for 3.3 Hz operation, which scales to ~ 9.4 cGy/min @ 1m for 10 Hz operation.

Accurate reconstruction of attenuation coefficients for CT critically depends on the stability of the X-ray beam profile. Figure 3B displays the centroid statistics of 1,000 X-ray shots overlaid on their elliptical average profile with FWHM divergences of 124 mrad (vertical) and 72 mrad (horizontal). The measured pointing fluctuations are 2.4 × 2.1 $mrad^2$ (r.m.s.) for single-shot profiles and further reduce to 0.4 × 0.3 $mrad^2$ for 100-shot accumulated profiles, which correspond to relative deviations of less than 3% and 0.5% of the overall divergence, respectively.

The bremsstrahlung spectrum was reconstructed using a spectrometer comprising 15-sector tungsten filters. Figure 3C shows the spectrum averaged from 100 shots, along with the attenuation profile obtained with the filters. The measured bremsstrahlung spectrum exhibits a peak at 1.9 MeV and a mean energy of 13.1 ± 0.9 MeV, with a high-energy tail extending beyond 100 MeV (primarily determined by the electron energy upper limit). With a half-value layer exceeding 20 mm in steel, this high-energy source is particularly well-suited for the NDT of large-scale, dense materials.

High-precision μCT demands a small source size. During the electromagnetic cascade of electron beams within a metal converter target, higher-energy electrons tend to retain their forward direction as they propagate, whereas lower-energy electrons scatter more strongly and spread over a wider region. As a result, the high-energy portion of the bremsstrahlung radiation originates from a more confined region, yielding a smaller source size. Compared to conventional bremsstrahlung sources driven by MeV-level RF linacs, the use of ~ 100 MeV electron beams in LWFA-driven sources significantly enhances the high-energy photon yield. This enhancement, coupled with the intrinsically micrometer-scale transverse size of the electron beam, facilitates a substantially smaller source size. We characterized the source size using penumbral imaging of a cylindrical tungsten-steel object at 15.5× magnification, deriving an FWHM source size of 38 μm from the measured edge blur (Fig. 3D). To assess the resolution performance, a 50-μm (10 lp/mm) line pair phantom was imaged at 6× magnification. The radiograph accumulated from 500 shots (Fig. 3E) clearly resolves the pattern with 33% contrast, far surpassing the discernible threshold (10% contrast). The intrinsic spatial resolution was estimated to be 35 μm at 10% contrast, assuming a Gaussian point spread function. As this result is constrained by our detector resolution (~ 150 μm), further developments in high-density, large-area polycrystalline transparent ceramic scintillators(*41*) are expected to enhance system resolution to better than 20 μm(*42*).

To demonstrate the μNDT capability for dense materials, we performed μCT on a CFM56 Stage 1 aeroengine turbine blade cast from a nickel-based superalloy, utilizing over 60,000 consecutive exposures at a repetition rate of 3.3 Hz. Designed to withstand extreme thermal loads, this component incorporates intricate internal geometries (specifically, serpentine cooling channels with turbulators) and a surface patterned with hundreds-of-micrometer film-cooling holes. Such blades present a stringent inspection challenge, as they are susceptible to both manufacturing anomalies (e.g., residual cores within internal channels) and operational defects (e.g., thermally induced cracks and voids) with characteristic dimensions typically below 100 μm.

A total of 360 projections were acquired at 1° angular intervals. Each projection, averaged over 150 consecutive exposures, yielded a signal-to-noise ratio (SNR) better than 100. Tomographic reconstructions were generated using the Feldkamp-Davis-Kress (FDK) algorithm with the Astra Toolbox(*43*). Figures 4A and 4B show 3D reconstructions of external and internal structures, revealing fine features such as film-cooling holes and internal channels. Supplementary movies provide the 3D volumetric rendering from multiple viewing angles (Supplementary Materials Movies S1 and S2). Figure 4C presents four representative transverse slices spanning the blade from tip to root (tenon). These cross-sections reveal the morphological evolution of the internal cooling channels, which taper from irregular, large-bore geometries at the tip to smaller, quasi-elliptical profiles near the root. To quantify the spatial resolution of the reconstruction, the modulation transfer function (MTF) was measured across the blade edge (Fig. 4D). A spatial resolution of 43 μm, corresponding to a spatial frequency of 11.6 $mm^{-1}$, was determined at an MTF of 10%. This performance surpasses the millimeter-level resolution typical of conventional RF linac-based systems and improves upon the sub-millimeter (~ 0.6 mm) capability recently reported for a 300-TW laser-driven MeV source(*44*).

Another critical demand in advanced manufacturing is μNDT of large-scale components composed of complex, heterogeneous materials. For instance, metallic spacecraft cabins and engine casings coated with composite layers are susceptible to interfacial debonding and fiber delamination, requiring a meter-scale FOV and 100-μm spatial resolution for defect inspection. To extend the effective FOV within a compact imaging distance, we implemented a four-image stitching strategy using two CsI-based detectors, expanding the lateral detection width to 430 mm (Fig. 5B). Combined with a turntable-offset scanning geometry, where the object rotation axis is offset by 170 mm, the scan diameter was increased to > 650 mm (Fig. 5A), enabling high-resolution volumetric imaging of large-scale objects.

To validate the penetration capability and spatial resolution for such samples, we imaged a 640-mm-diameter phantom consisting of an aluminum ring embedded with ceramic layers, serving as a surrogate for aeroengines and small launch vehicles. The tomogram presented in Fig. 5C was reconstructed from 1,440 projections using more than 170,000 X-ray pulses produced at 3.3 Hz. Owing to the contrast in X-ray attenuation coefficients between aluminum and the fibrous ceramic material, the two constituents are clearly distinguished. A variety of defects (both intrinsic and prefabricated) with characteristic dimensions on the 100-μm scale were resolved. These include interfacial debonding between the ceramic coating and the aluminum ring (Fig. 5C(i)), prefabricated slits between aluminum layers (Fig. 5C(ii)), and material inhomogeneities or impurities within the aluminum ring (Fig. 5C(iii)).

Conventional RF linac-based MeV NDT systems remain indispensable for most established industrial inspection tasks. However, the pursuit of higher detection precision, together with emerging manufacturing paradigms, is driving rapidly growing demand for high-energy μNDT capabilities. Our results establish LWFA technology as a compelling route toward high-resolution, high-energy radiation sources that extend beyond the performance envelope of conventional systems.

Scientific and industrial inspection scenarios vary widely in material composition, structural dimensions, and environmental conditions. The compact footprint and rapid on-site deployability of our system allow it to address such diverse and complex use cases. In contrast, clinical medical applications typically involve more standardized subjects and fixed installations within controlled environments. Operation in dedicated facilities—like those used for magnetic resonance imaging—would be expected to further enhance system stability and reliability, suggesting that LWFA-based sources could, in principle, meet the stringent requirements of future clinical translation.

As with many emerging technologies, successful deployment in industrial settings serves as a benchmark of maturity. Bridging the gap between laboratory plasma physics and real-world utility has been a longstanding objective for application-driven LWFA research. Compared with other LWFA-based modalities, such as X-ray phase-contrast imaging(*45*), X-ray absorption spectroscopy(*46*), and very-high-energy electron radiotherapy(*36*), the bremsstrahlung mode presents more challenges. Residual laser pulse damage and contamination from conversion targets impose tighter constraints on system design, shielding, and maintenance. Our results therefore not only unlock the domain of sub-50 μm 3D imaging with tens-of-MeV X-rays but also signify a pivotal step in the evolution of LWFA technology—marking its maturation from a laboratory-scale concept into a robust, field-deployable tool suitable for demanding real-world applications.

## Methods

**40 TW Ti:Sapphire laser.** The configuration of the 40 TW Ti:Sapphire laser system is shown in Fig. S3. A passively mode-locked Ti:Sapphire oscillator generates a continuous pulse train with a repetition rate of ~ 80 MHz and a spectral width of 104 nm (FWHM). The pulses are stretched to ~ 500 ps using an Offner-type aberration-free stretcher, then reduced to a 10 Hz repetition rate via a Pockels-cell injector inside a regenerative amplifier. The pulse energy is amplified to 15 mJ by the regenerative amplifier and further increased to 1.4 J by a six-pass amplifier. The r.m.s. energy stability over 10 hours was measured to be 0.63% for the regenerative amplifier (Fig. S3C) and 0.54% for the six-pass amplifier (Fig. S3D). A two-pass pulse cleaner, positioned between amplifier stages, enhances the temporal contrast to $>10^8$ on nanosecond timescales. After compression and focusing, the laser pulses with energy up to 1 J and durations of 25 fs (FWHM) are focused to a spot diameter of 10 μm (FWHM) by an f/9 off-axis parabolic mirror. For system miniaturization, key components (oscillator, amplifiers, pump lasers, Pockels cells, delay signal generators, and flexible optical frames) are self-designed and manufactured to high engineering standards. Each laser module includes an independent, high-precision temperature and humidity control system with real-time internal parameter monitoring. Heat-generating components such as power supplies and pump laser chillers are placed in an isolated room within the container enclosure to maintain stability.

**Electron beam diagnostics.** The electron beam profiles were measured by optically imaging a $Gd_2O_2S$ screen (DRZ PI-200, Mitsubishi Chemical) placed along the beam axis. The beam charge $Q$ was calculated as

$$Q = \frac{N}{\eta T \Omega S}$$

where $N$ is the signal count, $\eta$ is the photon-to-count conversion efficiency of the camera, $T$ is the optical transmission, $\Omega$ is the effective lens solid angle, and $S$ is the charge-to-photon conversion efficiency of the phosphor screen. Electron energy spectra were measured using a 120-mm-long permanent dipole magnet. The intensity distribution of the magnetic field was measured to establish the relationship between deflection distance and electron energy.

**Bremsstrahlung X-ray diagnostics.** The position of the tungsten target was optimized to maximize photon flux. Based on Monte Carlo simulations that balance photon flux and source size, a 3-mm tungsten target was selected. To prevent damage from residual laser pulses, a 100-μm polyethylene tape was mounted in front of the tungsten target. Bremsstrahlung X-rays were recorded using a shielded scintillator-based detector comprising a CsI scintillator screen, an optical transmission system, and an electron-multiplying charge-coupled device (EMCCD). X-ray spectra were reconstructed using an expectation maximization algorithm, based on X-ray attenuation profiles with a tungsten filter set (15 sectors, 0.5-20 mm thick). Radiation dose was measured using a calibrated dosimeter positioned 1 m from the source.

**Computed tomography.** CT of the turbine blade was performed at an imaging magnification of 2.3, with the detector placed 3.2 m from the source. The CsI-based detector had an effective view of $9.8 \times 9.8$ cm$^2$ and equivalent pixel size of 96 μm. Projections were acquired at 1° intervals over 360°, with each projection accumulated from 150 pulses at 3.3 Hz. The total scan duration was ~ 8 hours, including image acquisition, dark/flat field collection, stage motion, and detector readout.

For the aluminum-ceramic phantom, a turntable-offset CT configuration with a 170 mm axis displacement and 1.3× magnification was used, where the detectors were placed 3.2 m from the source. A total of 1,440 projections were acquired with a 0.25° angular interval. Each projection was stitched from four sub-images, with each sub-image averaged from 60 X-ray shots. Two CsI-based detectors recorded the sub-images simultaneously, each with an effective view of $12 \times 12$ cm$^2$ and equivalent pixel size of 117 μm. A single lateral translation of the detector produced four sub-images per projection angle.

For both CTs, a dark-frame image $D$ (X-ray off) was subtracted to correct for detector dark current. A flat-field image $G$ (object off), averaged over 1,000 shots, was used to correct for pixel-to-pixel response. Final projection images $P$ were then obtained as

$$P = -\ln\left(\frac{R - D}{G - D}\right)$$

where $R$ is the raw projection image.

**Acknowledgments:** The simulation work was supported by Center of High Performance Computing, Tsinghua University. The authors acknowledge Dr. Junjiang Li, Dr. Qing Ye, and Dr. Baihong Jiang for their assistance with CT reconstruction.

**Funding:**

Strategic Priority Research Program of the Chinese Academy of Sciences (XDB0530000)

IHEP Talent Introduction Program (E65152U1)

National Natural Science Foundation of China (12574380)

Discipline Construction Foundation of "Double World-class Project"

Science Fund Program for Distinguished Young Scholars of the National Natural Science Foundation of China (Overseas)

Key Scientific Research Projects of Henan Provincial Colleges and Universities (25ZX002)

Natural Science Foundation of Henan Province (252300421300)

**Author contributions:** W.L. conceived the idea and supervised the project. J.H. managed the project. W.L., J.H., and B.G. designed the prototype. B.G. and X.N. maintained the system. B.G., D.L., Y.M., W.Z., M.W. and S.W. conducted the experiments. B.G. and D.L. analyzed the data. B.G., Y.W. and W.L. drafted the manuscript. All authors contributed to the review of the manuscript.

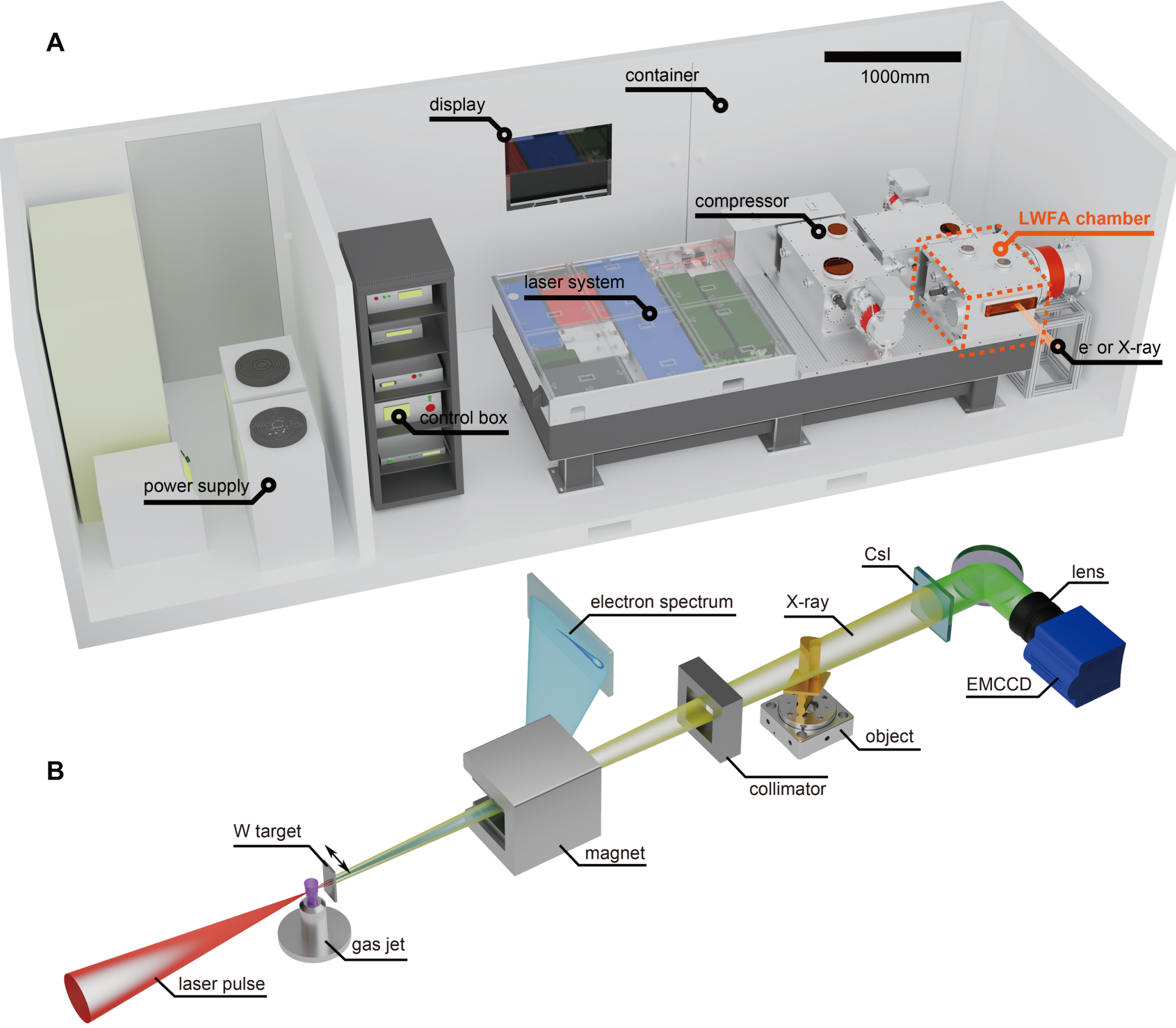

**Fig. 1. Configuration of the LWFA system.** (**A**) Schematic layout of the system. A compact LWFA driven by a 40 TW laser is integrated into a transportable container with a volume of ~ 50 m$^3$. Thermal management is implemented by separating heat-producing devices such as power supplies and chillers in an isolated room, minimizing their impact on the LWFA. All the modules are supported by a high-precision cooling system, while environmental parameters (e.g., temperature and humidity) and laser diagnostics (e.g., beam profile, spectrum, and energy) are continuously monitored to ensure long-term operational stability. (**B**) Schematic of the LWFA and the bremsstrahlung source. Laser pulses (red) with a 25-fs duration and a 10-µm focal spot size are focused onto a supersonic gas jet generated by a de Laval nozzle. In electron mode, electrons (blue) are deflected onto a $Gd_2O_2S$ scintillator screen for spectral measurement. In X-ray mode, electron beams strike a tungsten converter to produce bremsstrahlung X-rays (yellow). A polyethylene-tungsten collimator shields residual electrons deflected by a dipole magnet. After exiting the container, X-rays transmit through objects mounted on a rotation stage and are detected by a CsI-based detector.

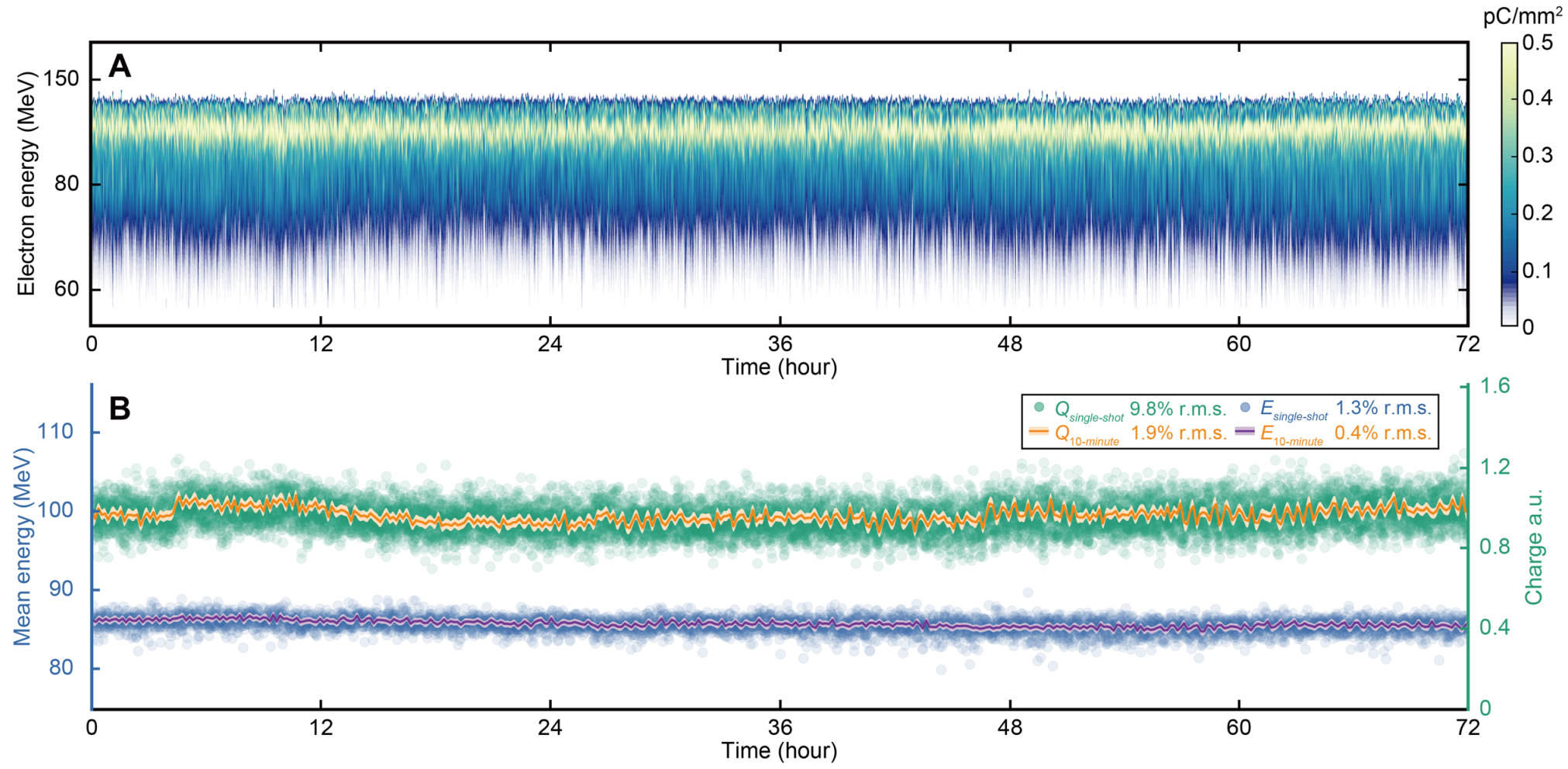

**Fig. 2. Long-term electron-beam stability.** (**A**) Electron spectra acquired continuously over 72 hours, driven by a laser pulse energy of 630 mJ (on target) and a plasma electron density of $1.2 \times 10^{19}$ cm$^{-3}$. The data were acquired at 0.1 Hz to reduce the data volume. (**B**) Statistics of mean energy and charge (>70 MeV) for the 72-hour period. The r.m.s. shot-to-shot fluctuations were 1.3% (86.2 ± 1.1 MeV) in mean energy ($E_{single\text{-}shot}$) and 9.8% (164 ± 16 pC) in charge ($Q_{single\text{-}shot}$). The r.m.s. fluctuations for 10-minute integrated data (corresponding to 60-shot averages) were reduced to 0.4% (mean energy, $E_{10\text{-}minute}$) and 1.9% (charge, $Q_{10\text{-}minute}$), respectively.

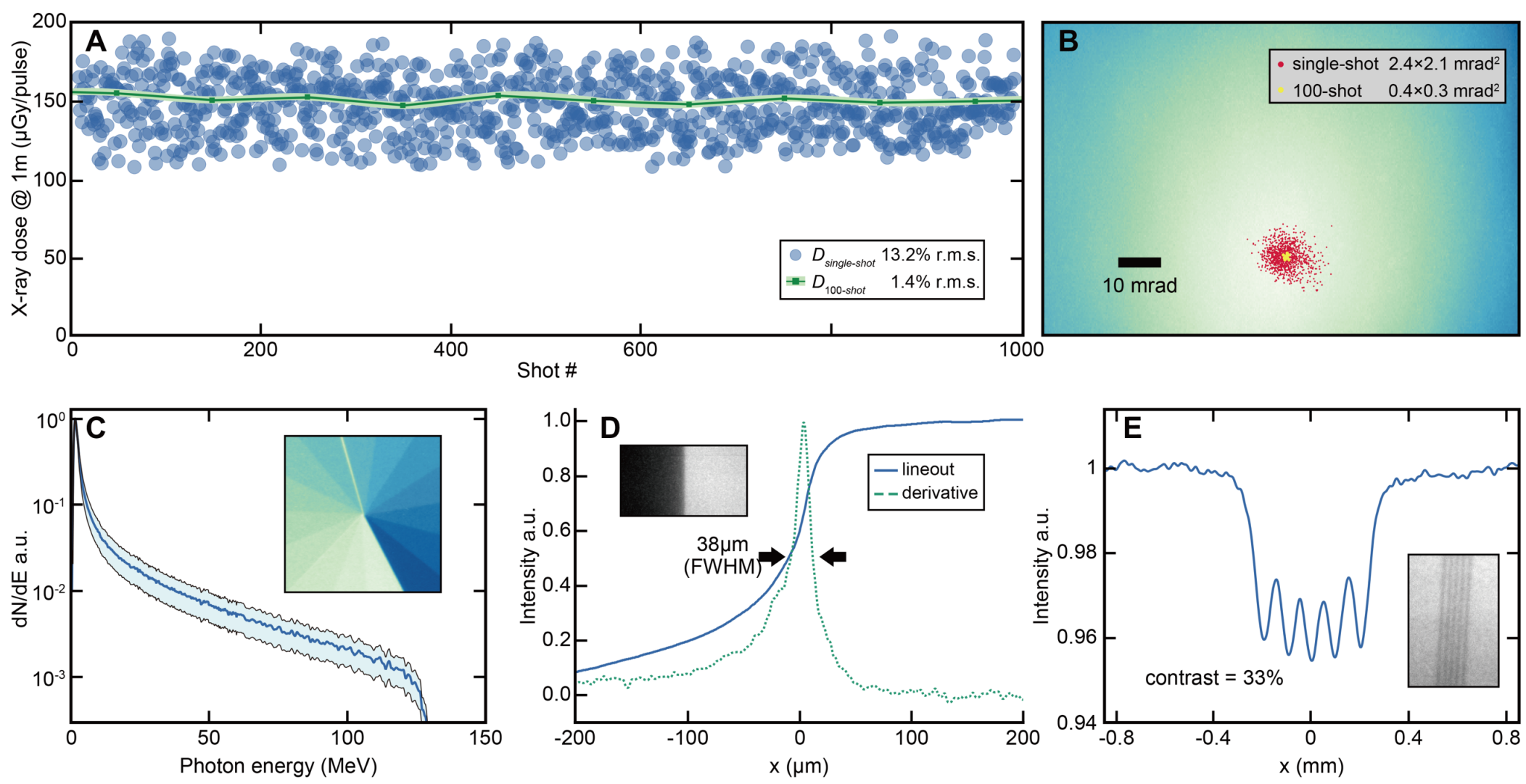

**Fig. 3. Characterization of bremsstrahlung X-rays.** (**A**) X-ray dose stability over 1,000 consecutive shots acquired at 3.3 Hz. The r.m.s. shot-to-shot fluctuation ($D_{single\text{-}shot}$) was 13.2% (157 ± 21 μGy/pulse). The fluctuation for 100-shot averages ($D_{100\text{-}shot}$) was reduced to 1.4%. (**B**) Average X-ray profile from 1,000 shots overlaid with the centroid positions for single-shot profiles (red points) and 100-shot accumulated profiles (yellow points). The average X-ray divergence was 124 × 72 mrad$^2$ (FWHM), with r.m.s. pointing stabilities of 2.4 × 2.1 mrad$^2$ (single-shot) and 0.4 × 0.3 mrad$^2$ (100-shot). (**C**) Average bremsstrahlung X-ray spectrum derived from 100 consecutive shots, peaking at 1.9 MeV. The mean energy was 13.1 ± 0.9 MeV. The inset shows the average X-ray profile attenuated by a tungsten filter set. (**D**) Penumbral edge intensity lineout (blue solid) and its first-order derivative (green dashed) obtained from the image of a tungsten steel cylinder. The inset presents a 400-shot averaged radiograph at 15.5× magnification. The accumulated source size was 38 μm (FWHM). (**E**) Intensity lineout extracted from a 500-shot averaged radiograph of a 10 lp/mm line pair phantom at 6× magnification. The 50-μm patterns are clearly resolved with 33% contrast, significantly exceeding the discernible threshold (10% contrast).

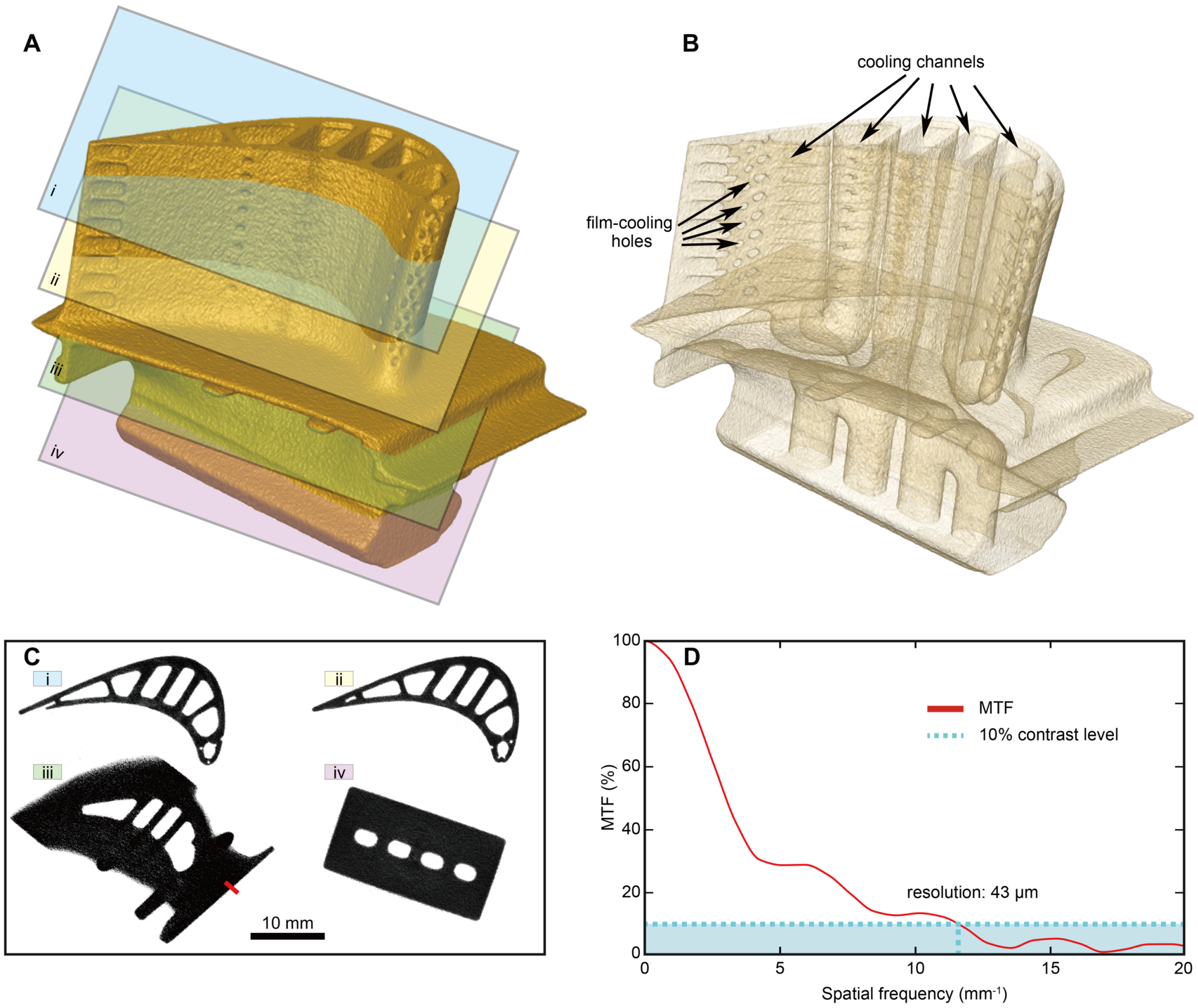

**Fig. 4. μCT of an aeroengine turbine blade.** (**A**, **B**) 3D reconstructions of the outer contour and internal structures of an aeroengine turbine blade. The hundreds-of-micrometer film-cooling holes on the surface and intricate internal cooling channels are both clearly visible. (**C**) Reconstructed transverse slices of the turbine blade at the positions marked in (A). (**D**) MTF obtained from a lineout across the blade edge in (C). The 10% contrast threshold (cyan dashed line) corresponds to a spatial resolution of 43 μm (11.6 lp/mm).

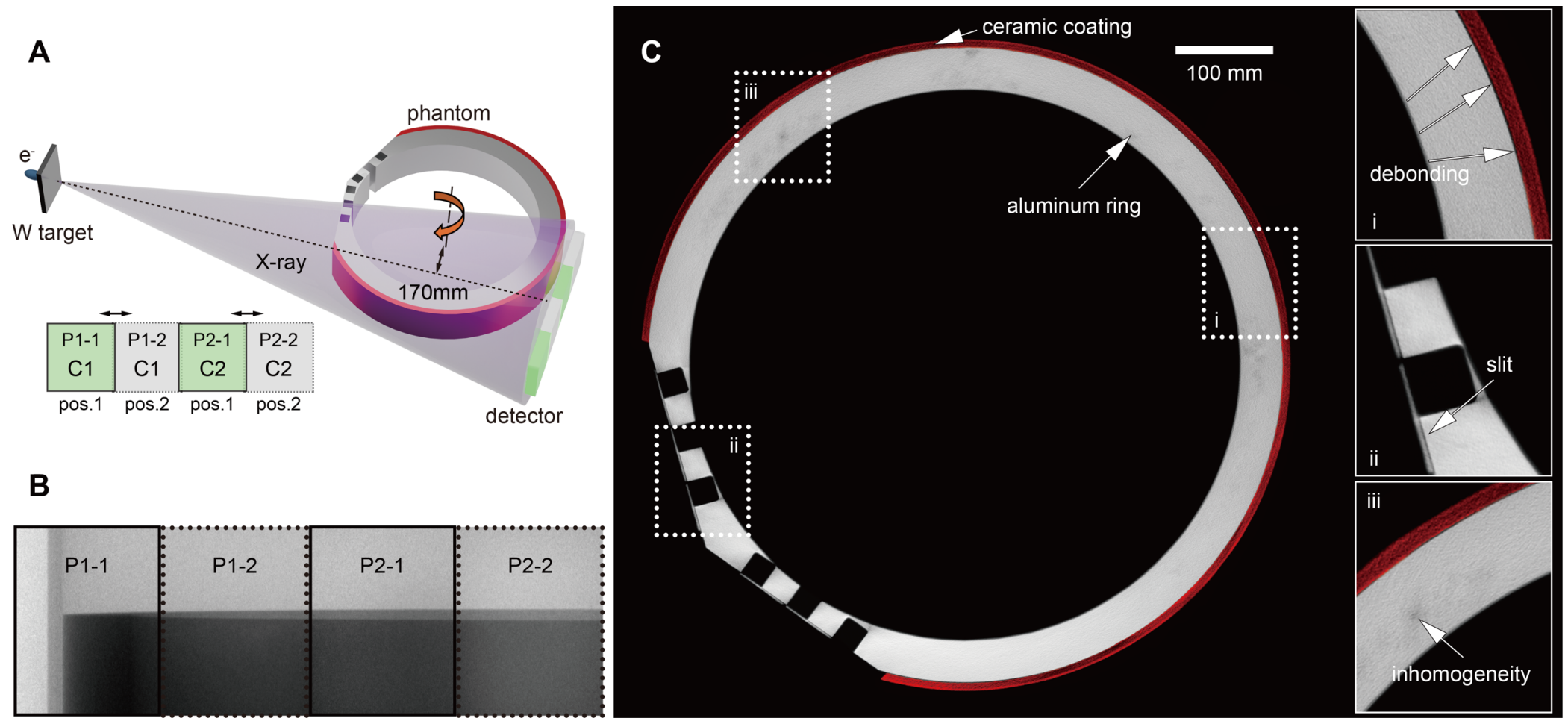

**Fig. 5. Large-FOV µCT.** (**A**) Schematic of the large-FOV CT scan and image stitching geometry. The object's rotational axis was offset by 170 mm from the center. Two CsI-based detectors (C1 and C2), spaced 105 mm apart, were sequentially moved to two positions (pos.1 and pos.2) with a 100 mm separation via a motorized translation stage. This geometry produced four sub-projections: P1-1 (pos. 1) and P1-2 (pos. 2) from C1, and P2-1 (pos. 1) and P2-2 (pos. 2) from C2. These four sub-projections were computationally stitched to form a single, complete projection for the final CT reconstruction. (**B**) A representative stitched projection obtained from 120 exposures, with each sub-projection averaged from 60 exposures. (**C**) A representative reconstructed slice of a 640 mm aluminum-ceramic phantom imaged with >170,000 X-ray pulses (1,440 projections). The aluminum and ceramic layers are clearly distinguished. Internal 100-µm-scale defects (insets), including debonding between the layers (i), prefabricated slits (ii), and inhomogeneities (iii), are fully resolved.

# Supplementary Materials for

## Field deployment of a laser wakefield accelerator for on-site application

Bo Guo, Xiaonan Ning, Dexiang Liu, Yue Ma, Weiwang Zeng, Mingyuan Wei, Shengtai Wei, Jianfei Hua, Yang Wan, Wei Lu

Corresponding author: weilu@ihep.ac.cn

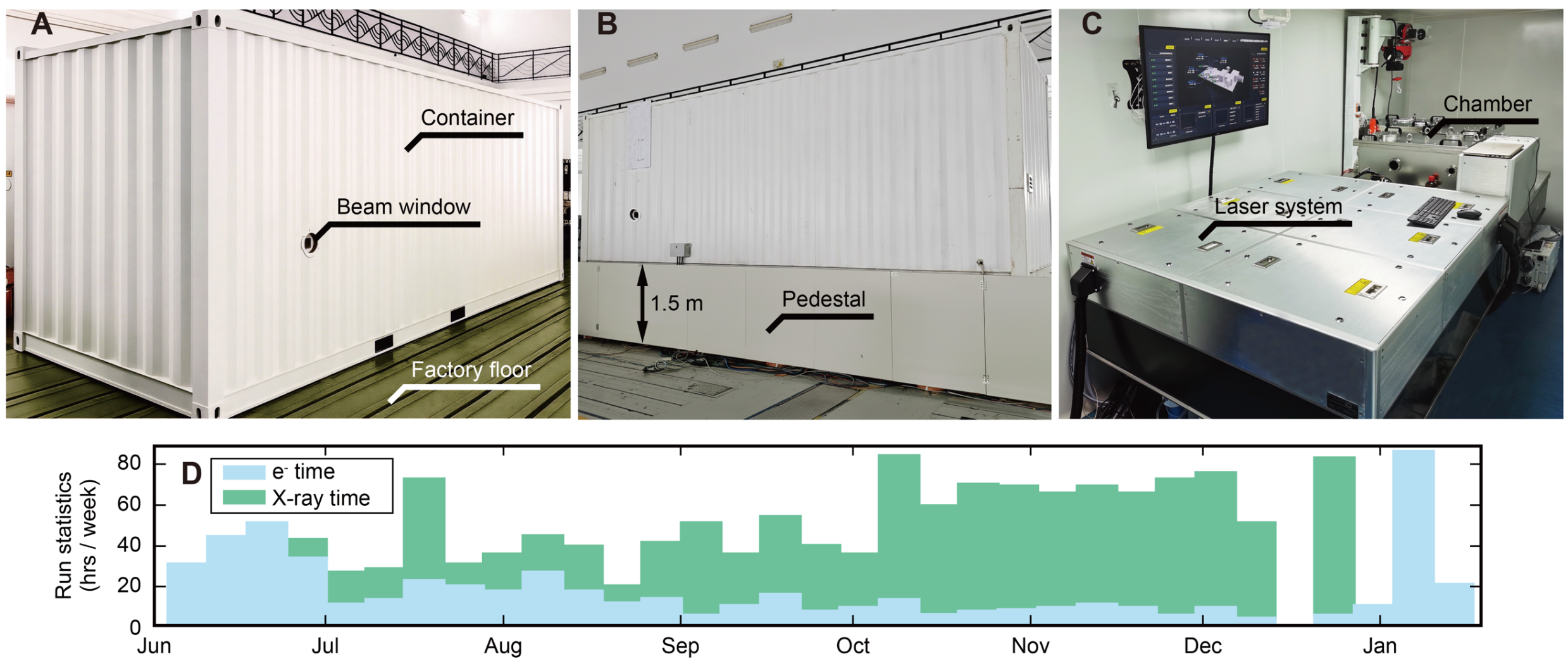

**Fig. S1. Photographs and operational statistics of the LWFA system.** (**A, B**) Two operational configurations within a radiation-shielded NDT factory adapting to different application needs. Switching between these configurations requires simple optical alignment with a downtime of a few days. (**C**) Interior view of the container. (**D**) Statistical summary of weekly full-power operation time over a 7-month period. The source operated on 151 days, with full-power operation at 3.3 Hz for more than 10 hours per day.

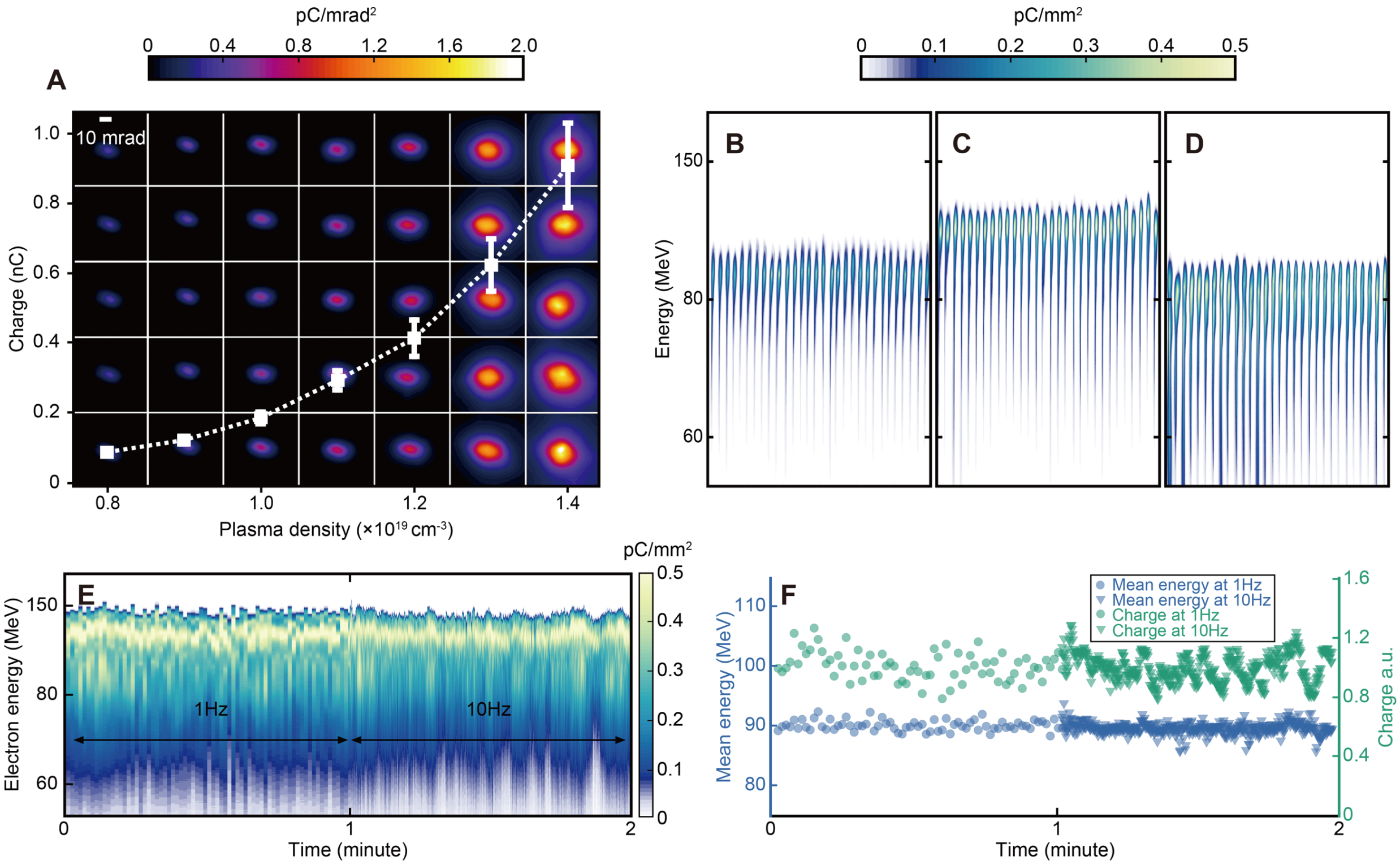

**Fig. S2. Characterization of electron beam.** (**A**) Charge statistics for 30 consecutive electron beams at plasma densities of $n_e$ = (0.8 – 1.4) × $10^{19}$ cm$^{-3}$, together with five consecutive profiles. The FWHM divergences ranged from (10.9 ± 0.9) × (8.3 ± 0.3) mrad$^2$ to (23.1 ± 1.2) × (17.4 ± 1.8) mrad$^2$, with centroid position deviations of <2 mrad (r.m.s.). (**B – D**) Spectral images of 30 consecutive shots at plasma densities of (0.9 – 1.3) × $10^{19}$ cm$^{-3}$. The mean energy was 82.7 ± 1.1 MeV (9 × $10^{18}$ cm$^{-3}$), 93.4 ± 1.4 MeV (1.1 × $10^{19}$ cm$^{-3}$), and 74.8 ± 2.1 MeV (1.1 × $10^{19}$ cm$^{-3}$). (**E**) Spectral images of electron beams acquired at 1 Hz and 10 Hz over a continuous one-minute interval. (**F**) Shot-to-shot statistics of mean energy (blue markers) and beam charge (green markers) at 1 Hz (circles) and 10 Hz (triangles). The mean energy was 90.1 ± 0.9 MeV (1 Hz) and 89.6 ± 0.9 MeV (10 Hz), corresponding to r.m.s. fluctuations of ~ 1%, while the r.m.s. charge (>70 MeV) fluctuations were 9.6% (1 Hz) and 9.4% (10Hz).

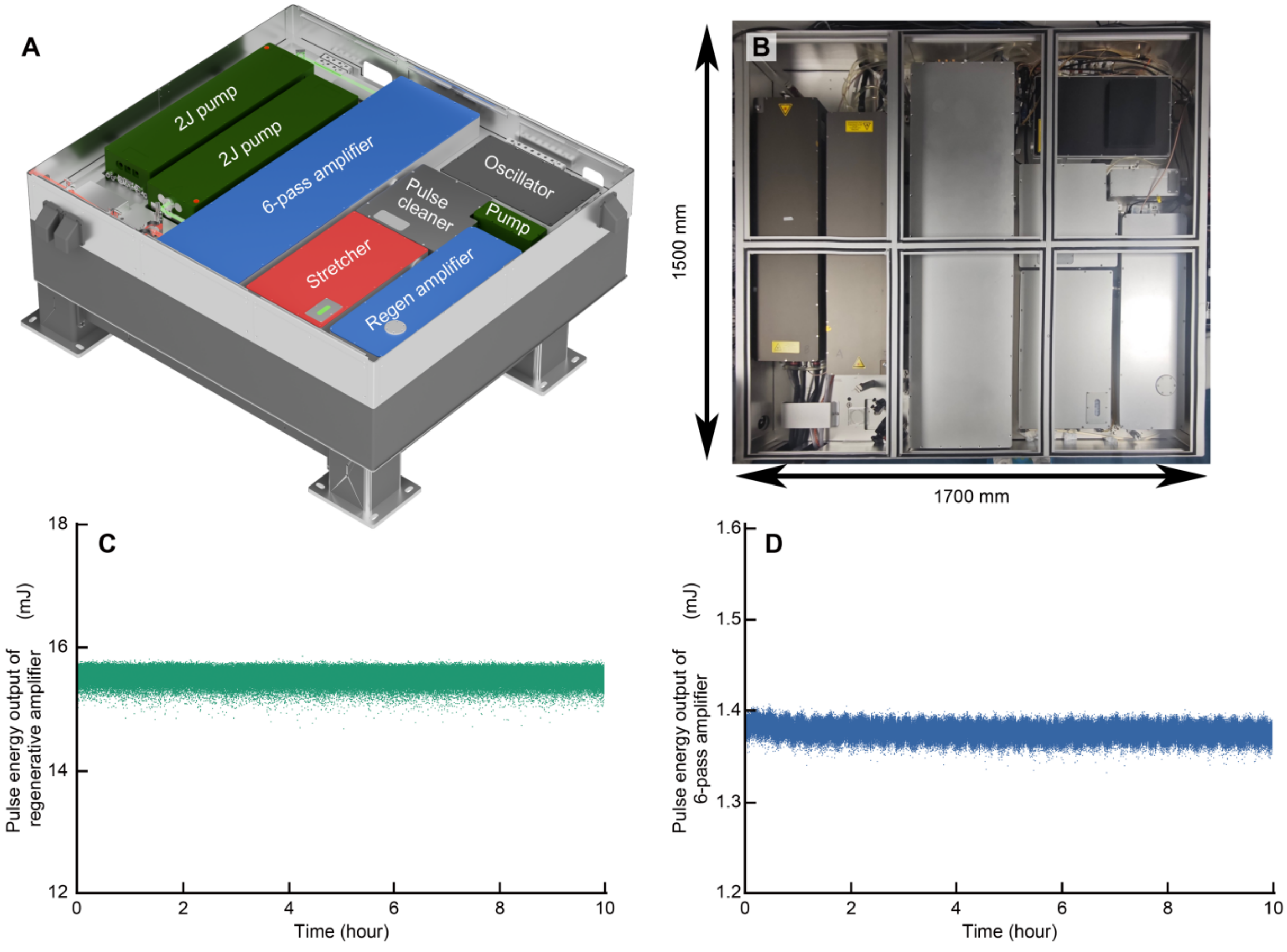

**Fig. S3. Construction and stability of the 40 TW Ti:Sapphire laser.** (**A**) Schematic layout of the laser system. (**B**) Photograph of the laser system. (**C**) Shot-to-shot output pulse energy of the regenerative amplifier over a consecutive 10-hour period, showing an r.m.s. stability of 0.63%. (**D**) Shot-to-shot output pulse energy of the six-pass amplifier over a consecutive 10-hour period, with an r.m.s. stability of 0.54%.